\documentclass[osajnl,twocolumn,showpacs,superscriptaddress,11pt]{revtex4-1} 
\usepackage{amsmath,amssymb,graphicx}

\begin{document}
\title{Large-aperture wide-bandwidth anti-reflection-coated silicon lenses for millimeter wavelengths}

\author{R. Datta}\email{Corresponding author: dattar@umich.edu}
\author{C. D. Munson}
\affiliation{University of Michigan Physics Department, 450 Church St., Ann Arbor, MI 48109}

\author{M. D. Niemack}
\affiliation{Cornell University Physics Department, 109 Clark Hall, Ithaca, New York 14853}
\affiliation{National Institute of Standards and Technology, 325 Broadway, Boulder, CO 80305}

\author{J. J. McMahon}
\affiliation{University of Michigan Physics Department, 450 Church St., Ann Arbor, MI 48109}

\author{J. Britton}
\affiliation{National Institute of Standards and Technology, 325 Broadway, Boulder, CO 80305}

\author{E. J. Wollack}
\affiliation{NASA Goddard Space Flight Center, 8800 Greenbelt Road, Greenbelt, MD 20771}

\author{J. Beall}
\affiliation{National Institute of Standards and Technology, 325 Broadway, Boulder, CO 80305}

\author{M. J. Devlin}
\affiliation{University of Pennsylvania Department of Physics and Astronomy, 209 South 33rd St., Philadelphia, PA 19104}

\author{J. Fowler}
\affiliation{National Institute of Standards and Technology, 325 Broadway, Boulder, CO 80305}

\author{P. Gallardo}
\affiliation{Departamento de Astronomia y Astrofisica, Facultad de Fisica, Pontificia Universidad Catolica, Casilla 306, Santiago 22, Chile}

\author{J. Hubmayr}
\author{K. Irwin}
\affiliation{National Institute of Standards and Technology, 325 Broadway, Boulder, CO 80305}

\author{L. Newburgh}
\affiliation{Princeton University Department of Physics, Jadwin Hall, P. O. Box 708, Princeton, NJ 08544}

\author{J. P. Nibarger}
\affiliation{Boulder Micro-Fabrication Facility, National Institute of Standards and Technology, 325 Broadway, MS 817.03, Boulder, CO 80305}

\author{L. Page}
\affiliation{Princeton University Department of Physics, Jadwin Hall, P. O. Box 708, Princeton, NJ 08544}

\author{M. A. Quijada}
\affiliation{NASA Goddard Space Flight Center, 8800 Greenbelt Road, Greenbelt, MD 20771}

\author{B. L. Schmitt}
\affiliation{University of Pennsylvania Department of Physics and Astronomy, 209 South 33rd St., Philadelphia, PA 19104}

\author{S. T. Staggs}
\affiliation{Princeton University Department of Physics, Jadwin Hall, P. O. Box 708, Princeton, NJ 08544}

\author{R. Thornton}
\affiliation{West Chester University of Pennsylvania Department of Physics, 700 South High St., West Chester, PA 19383}

\author{L. Zhang}
\affiliation{Princeton University Department of Physics, Jadwin Hall, P. O. Box 708, Princeton, NJ 08544}

\begin{abstract}

The increasing scale of cryogenic detector arrays for sub-millimeter and millimeter wavelength astrophysics has led to the need for large aperture, high index of refraction, low loss, cryogenic refracting optics. Silicon with $n = 3.4$, low loss, and relatively high thermal conductivity is a nearly optimal material for these purposes, but requires an antireflection (AR) coating with broad bandwidth, low loss, low reflectance, and a matched coefficient of thermal expansion. We present an AR coating for curved silicon optics comprised of subwavelength features cut into the lens surface with a custom three axis silicon dicing saw. These features constitute a metamaterial that behaves as a simple dielectric coating. We have fabricated  and coated silicon lenses as large as 33.4 cm in diameter with coatings optimized for use between 125-165 GHz.  Our design reduces average reflections to a few tenths of a percent for angles of incidence up to $30^\circ$ with low cross-polarization. We describe the design, tolerance, manufacture, and measurements of these coatings and present measurements of the optical properties of silicon at millimeter wavelengths at cryogenic and room temperatures. This coating and lens fabrication approach is applicable from centimeter to sub-millimeter wavelengths and can be used to fabricate coatings with greater than octave bandwidth.

\end{abstract}

\maketitle 
\section{Introduction}

The development of large format superconducting detector arrays for millimeter and sub-millimeter astrophysics (e.g. \cite{Bintley:2012,Niemack:2008,Padin:2008}) has driven the need for high-throughput optical designs that maintain diffraction limited performance across the arrays (see \cite{Hanany:2012} for a review). Silicon is an excellent material for these applications due to its high index of refraction ($n = 3.4$), low loss-tangent (tan $\delta < 7 \times 10^{-5} $), and a relatively high thermal conductivity ($k$ $\sim$ 200 $W m^{-1} K^{-1}$ at 4 $K$ \cite{Thompson:1961}, which is only a factor of 50 lesser than high quality OFHC copper \cite{Sciver:2012} and orders of magnitude higher than plastic). The essential development required to realize optical designs using silicon optics at millimeter wavelengths is an appropriate anti-reflection (AR) coating to mitigate the 30\% (-5 dB) reflective loss from each optical surface.

AR coatings consist of one or more dielectric layers placed on the surface of refractive optical elements.  The index of refraction and thickness of the AR layers are chosen such that the reflections from the vacuum-AR interface and from the AR-substrate interface interfere and cancel.  If the optical path length through a single layer AR coating is one quarter wavelength, the refractive index of the coating is $\sqrt n$ (where $n$ is the index of the substrate), and the coating is free from dielectric losses, the cancelation is perfect at one particular frequency.  Applied to silicon, such quarter-wave coatings reduce reflections below $1\%$ over 1.25:1 bandwidth at normal incidence. 

Wider bandwidths and larger ranges of angles of incidence can be accommodated by adding additional layers to the AR coating to form a multi-layer coating.  These coatings require layers with a number of different refractive indices.  Cryogenic applications require that the coefficient of thermal expansion of the AR layers also be sufficiently matched to that of the lens substrate to prevent damage upon cooling.  The dielectric loss and birefringence of the AR coating material must also be carefully controlled.  Locating materials with these properties poses a significant challenge to implementing cryogenic wide band AR coatings for silicon. One avenue to solve this problem is to engineer materials with the required dielectric constant by cutting sub-wavelength features into the lens surface \cite{Collin:1990, Smith:2006}. Engineered optical materials with properties determined by their detailed geometric shape, size and orientation are referred to as artificial dielectrics or metamaterials.

Simulated dielectric metamaterial AR coatings consist of layers of sub-wavelength holes \cite{Kildal:1984}, posts, or grooves  \cite{Cohn:1961} cut into the substrate to be coated.  This approach allows the index of the AR coating to be precisely tuned by adjusting the geometry of the machined features, naturally solves the thermal expansion problem, and has loss lower than that of the substrate.  At cryogenic temperatures silicon has a dielectric loss tangent $\sim$ 100 times lower than typical plastics like Cirlex, which have previously found use as single layer AR coatings \cite{Lau:2006}. The advantages of  silicon metamaterial AR coatings include precisely controlled indices of refraction, inherently matched thermal expansion, and significantly reduced dielectric losses.

This approach has been successfully applied to plastic millimeter wave optics through direct machining and silicon optics for visible wavelengths through patterning of geometric structures which perform as a AR coating similar to that found on a moth's eye \cite{Thornton,Chapham,Motamedi}.  At millimeter wavelengths Zhang et al. \cite{Zhang:2009} fabricated an alternative class of simulated dielectric AR for silicon composed of metal resonant structures supported by a plastic substrate that had higher losses than the coating described here.   Schuster et al. \cite{Schuster:2005} fabricated single layer silicon micro-machined artificial dielectrics and Han et al. \cite{Han:2010} have used silicon immersion grating technology to fabricate planar artificial dielectrics for THz radiation.  Our work is the first demonstration of a broad-bandwidth metamaterial AR coating on silicon optics with finite curvature for millimeter wavelengths. 

The AR coated lenses presented here have been developed for the ACTPol \cite{Niemack:2010} project, a polarization sensitive receiver for the Atacama Cosmology Telescope (ACT, \cite{Fowler:2007}).  In Section \ref{sec:req} we present the design requirements for the ACTPol optics as they apply to this work. The design of the coating is described in Section \ref{sec:des}. In Section \ref{sec:fab} we describe the ACTPol lens design.  Cryogenic measurements of the dielectric properties of silicon samples taken from 45 cm diameter ingots at millimeter wavelengths are presented in Section \ref{sec:siliconprops}. In Section \ref{sec:descriptionofdicingsystem} we describe the fabrication process of the coating on the lenses and present reflection measurements of a finished lens. We conclude with a brief discussion of the range of applicability of this coating technique.

\section{Requirements}
\label{sec:req}

\begin{figure}[t]
\centerline{\includegraphics[width=0.5\textwidth, keepaspectratio]{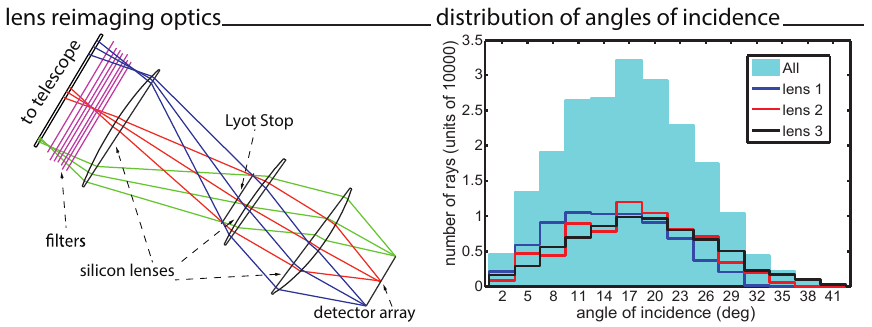}}
\caption{\footnotesize {\it Left:} A ray diagram of one set of ACTPol re-imaging optics, which includes three silicon lenses feeding a detector array.  {\it Right:} A histogram of the angles of incidence of rays at the surfaces of these three silicon lenses.}
\label{fig:actopt}
\end{figure}

The performance requirements for the AR coated silicon lenses described in this work are dictated by the scientific goals of the ACTPol project \cite{Niemack:2010}. This instrument is designed to detect the faint polarized signals from the cosmic microwave background using three detector arrays fed by independent re-imaging optics.  Two of the arrays are horn coupled polarimeters  for the 150 GHz band (passband from 125 to 165 GHz), while the third array will be a multichroic array including both 150 GHz and 90 GHz bands (`90 GHz' passband from 80 to 110 GHz) \cite{McMahon:2012}.  Extensive design studies demonstrated high index of refraction ($n\gtrsim 3$ ) lenses were needed to achieve high optical quality across the required field of view.  

In this paper we focus on the coatings for the 150 GHz band lenses. These coatings are conceptually similar to those which will be used for the broader-band multichroic lenses.  The bandwidth requirement for the 150 GHz band coatings is to have $> 99\%$ transmission between 125 and 165 GHz. The left panel of Figure~\ref{fig:actopt} shows the re-imaging optics for a 150 GHz array.  The optics include three plano-convex silicon lenses with diameters up to 33.4 cm.  The figure shows that rays passing through the optical system refract over a wide range of angles of incidence.  This is quantified in the right hand panel of Figure~\ref{fig:actopt} which shows that the distribution of angles of incidence is centered near $17^\circ$ and that more than 96\% of the rays have angles of incidence  $< 30^\circ$.  This sets the requirement that the coatings must be optimized to minimize reflections for angles of incidence between $0^\circ$ and $30^\circ$.

Since ACTPol is a polarization sensitive experiment, low cross-polarization is another requirement.  Studies of polarization systematics (e.g., \cite{Shimon:2008}) suggest that the CMB temperature to polarization leakage must be controlled to better than $1\%$ which  corresponds to a requirement that differences in the transmission for the two polarizations be 0.5\%.     

To reduce the thermal emission, all optics are cooled to 4K or below. Thus the AR coating must be able to withstand cryogenic cycling.

As described in \S \ref{sec:siliconprops} high resistivity silicon which is available in boules up to 45 cm in diameter, offers a cryogenic loss tangent $\tan \delta < 7 \times 10^{-5} $, an index of refraction of $n = 3.4$, and a relatively high thermal conductivity making it an ideal material for this optical design.  Our approach of directly machining metamaterial AR coatings into the lens surfaces guarantees that the coatings have low dielectric losses and coefficients of thermal expansion that are inherently matched to that of the silicon lenses.

\begin{figure}[t]
\centerline{\includegraphics[width=0.43\textwidth, keepaspectratio]{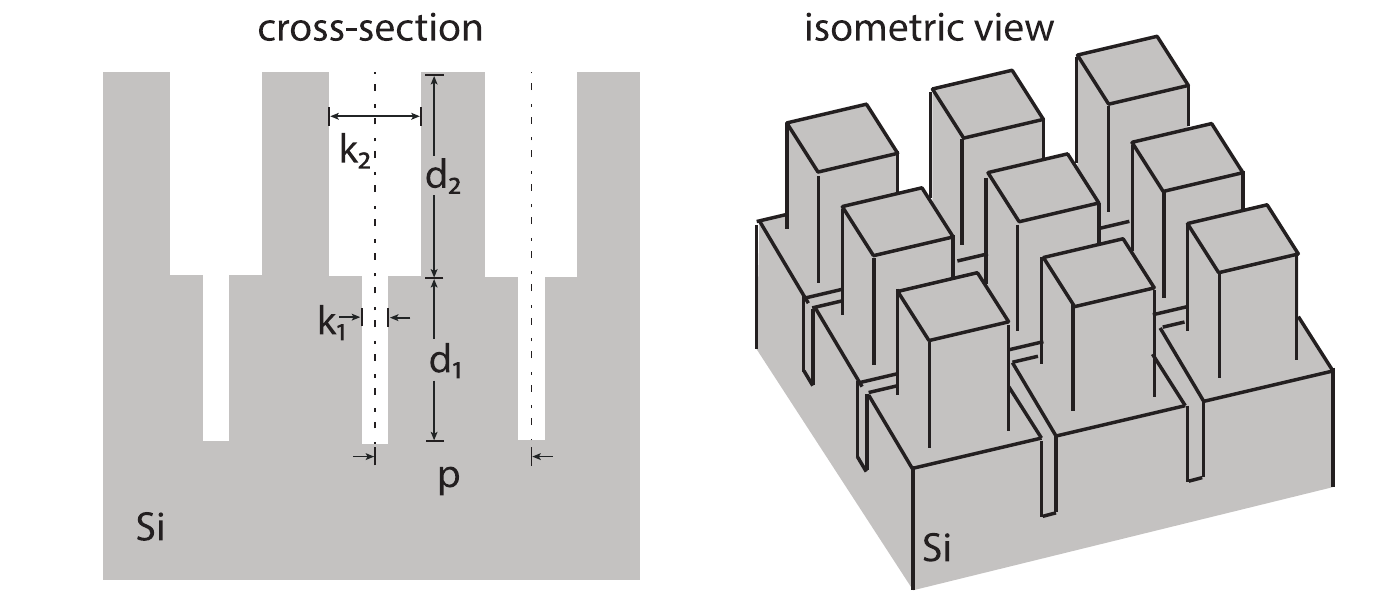}}
\caption{\footnotesize The geometry of the two layer metamaterial AR coating.  The left side shows a cross section through the center of the pillars indicating the design parameters discussed in the text.  The right side shows an isometric view of the structure resulting from making cuts along two perpendicular directions.}
\label{fig:geom}
\end{figure}

\section{AR Coating Design}
\label{sec:des}

The AR coating is comprised of two metamaterial layers.  Each layer consists of an array of square pillars on a square grid cut using silicon dicing saw blades by making two sets of parallel trenches rotated $90^\circ$ relative to each other.  It has been shown that any electromagnetic structure with an axis of n-fold (n$>$2) rotational symmetry must have polarization independent reflectance and zero cross-polarization when light is incident along this axis \cite{Mackay:1989}. Hence the $90^\circ$ rotational symmetry of this coating leads to zero cross-polarization at normal incidence and low birefringence at oblique incident angles. Figure~\ref{fig:geom} shows a cross-section and isometric view of this geometry which is parameterized by the depths $d_1$, $d_2$, and kerf widths $k_1$, $k_2$ of the inner and outer layer respectively and a pitch $p$. The outer layer is cut by using a blade of width $k_2$ and the inner layer is cut by making a second set of cuts to greater depth with a thinner blade of width $k_1$. We introduce the volume fill factor $v_f = (1-k/p)^2$ to facilitate comparison to published analytic calculations based on the second order effective medium theory developed by Rytov \cite{Rytov:1956} for the effective dielectric constant in the quasi-static limit ($p << \lambda$) \cite{Motamedi,Biber:2003,Brundrett:1994,Gaylord:1986}.

The metamaterial layers can be treated as a volume distribution of small electromagnetic scatterers characterized by electric and magnetic polarizability densities \cite{Collin:1990}. When the pitch is small compared to shortest wavelength of interest, the fields in the layer are homogenous \cite{Smith:2006} and one can define an effective dielectric constant which can be used to parameterize the propagation properties of the media. This effective dielectric function for the layer, $\epsilon_r^{eff}$, is a function of the density of scatterers in the layer \cite{Aspnes2:1982,Aspnes1:1982}, parametrized by the volume fill factor. As the feature size of a composite media approaches a significant fraction of the radiation wavelength the effective dielectric function becomes frequency dependent \cite{Egan:1982}. As the wavelength is further reduced the artificial dielectric structure no longer appears homogenous and these simple quasi-static considerations must be augmented to adequately model its behavior. In going beyond the quasi-static limit resonant effects, diffraction and scattering can occur in the artificial dielectric structure \cite{Kildal:1984,Gentner:2006,Raguin:1993,Matthaei}. The coatings described here, which operate between the quasi-static and diffractive regimes, rely on numerical simulations to estimate the relation between the volume fill factor and the effective refractive index of the metamaterial layer.  

The values of $d_1$, $d_2$, $k_1$, $k_2$, and $p$  were chosen using a three step design process.  A preliminary design was carried out using a classical analytic model consisting of sheets of dielectric material each with a constant thickness and index of refraction \cite{Jackson:1998}. These were varied to minimize the reflectance across the band yielding targets for the electrical thickness and index of refraction of our two layer coating.

\begin{figure}[t]
\centerline{\includegraphics{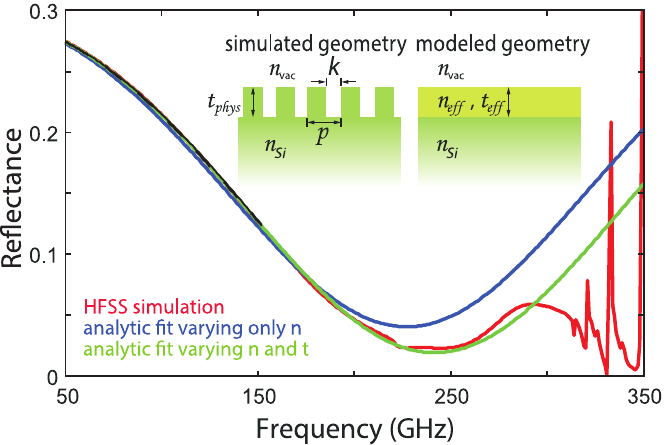}}
\caption{\footnotesize A comparison of the reflectance calculated from an HFSS numerical simulation of a pillar geometry to fits based on a model of a simple dielectric layer. The geometry is specified in the inset in the top. In the simulated geometry, $p$ = 400 $\mu$m, $t_{phys}$ = 220 $\mu$m, $k$ = 40 $\mu$m and $n_{Si}$ = 3.38. A comparison between fits where only the index $n$ is free and where both the index  $n$ and thickness $t$ are varied is shown. The HFFS simulation and the best fit curve differ by 5\% at 220 GHz ($\sim 1.36$ mm).This corresponds to the minimum wavelength, $\lambda_{5\%}$ (and a corresponding maximum frequency $f_{5\%}$) for the specified pitch, $p$ below (above) which the coating no longer behaves as a simple dielectric layer.} 
\label{fig:designcurves1}
\end{figure}

The second step was to translate the index and thickness from the simple model into the pitch, kerfs and depths for our coating geometry. This step required knowledge of the relation between the effective index of refraction of an array of square pillars and the pitch and kerf of that layer.  These relations were determined by fitting analytic models to HFSS (High Frequency Structure Simulator,  \cite{Ansoft:HFSS}) simulations of single layer coatings with a wide range of volume fill factors for a few different choices of pitch. The HFSS simulations were carried out using Floquet ports and master-slave boundary conditions to model a box containing a single post of the AR coating as an infinite periodic two dimensional array of features \cite{Ansoft:HFSS}.

\begin{figure*}[t]
\centerline{\includegraphics{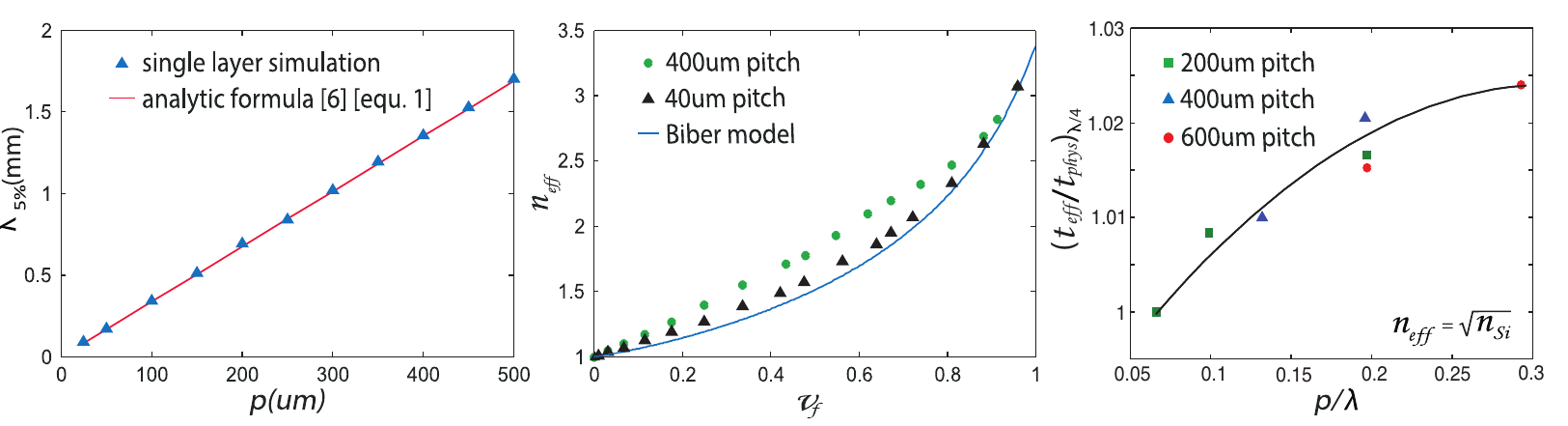}}
\caption{\footnotesize A summary of results from HFSS simulations of single layers of pillars with varying geometry. {\it Left:} The minimum wavelength $\lambda_{5\%}$ for which the single pillar layer is fit by an analytic model to 5\% accuracy as a function of pitch $p$ at normal incidence. {\it Center:} The effective index of refraction $n_{eff}$ as a function of the volume fill factor $v_f$ obtained by fitting the simulated reflectances upto the minimum wavelength $\lambda_{5\%}$ = 1360 $\mu$m for a pitch,  $p$ = 400 $\mu$m, 40 $\mu$m. The line labeled ``Biber" is a prediction from the analytic Biber model \cite{Biber:2003}, which corresponds to the quasi-static circuit approximation for the media.  {\it Right:} The thickness $t_{eff}$ for a slab dielectric quarter-wave coating divided by the thickness $t_{phys}$ of pillars forming an effective quarter wave coating as a function of $p/\lambda$. The index of the simulated dielectric coating was held fixed at $\sqrt {n_{Si}}$ for these simulations. A quadratic fit is plotted to guide the eye.}
\label{fig:designcurves2}
\end{figure*} 

Figure \ref{fig:designcurves1} shows a representative simulation of a single layer pillar geometry with insets showing the simulated geometry and the simple dielectric layer modeled in the fit.  The fit constrained the effective index $n_{eff}$, the effective electrical  thickness $t_{eff}$ and the maximum frequency to which our metamaterial coating behaves as a simple dielectric.  For the purposes of our fit we define this maximum frequency as the point where the simulated reflectance and the analytic fit disagree at more than 5\% absolute reflectance.  We refer to this frequency as $f_{5\%}$ and the corresponding wavelength $\lambda_{5\%}$.  This indicates the transition between specular and diffractive behavior of the coating. 
 
A second fit which fixed the electrical thickness to the physical thickness of the pillars is also shown in Figure \ref{fig:designcurves1}.  This fit implies a lower $f_{5\%}$.  The discrepancy between the best fit thickness $t_{eff}$ and the physical thickness $t_{phys}$ of the metamaterial layer arises since the geometric structure of the pillars results in fringing of fields at the junction between layers causing a perturbative shift between the position of the physical interface and the effective location of the junction's electrical reference plane.  An analogous effect is encountered in the design of metallic waveguide structures \cite{Matthaei}, which can be analytically treated. 

Figure \ref{fig:designcurves2} shows the results from fitting a number of HFSS simulations with different pitch and fill factor to the analytic model.  Plots include the breakdown wavelength $\lambda_{5\%}$, the effective index $n_{eff}$, and the electrical thickness of a layer with index $\sqrt{n_{Si}}$ expressed as a ratio of the thickness of a quarter wave homogenous dielectric layer and a quarter wave metamaterial layer.   

The recovered $\lambda_{5\%}$ is in good agreement with the analytic condition derived for a silicon optic in vacuum \cite{Kildal:1984, Raguin:1993} to prevent diffraction from a grating array,
\begin{equation}
	\ p < \frac{\lambda}{(n_{Si} + \sin\theta_i)},
\label{eq:Kildal}
\end{equation}
 where $p$ is the pitch, $\lambda$ is the wavelength, $n_{Si}$ is the index of silicon, and $\theta_i$ is the angle of incidence.   This analytic expression shows that the pitch must be smaller than implied by the $\lambda_{5\%}$ calculated at normal incidence to minimize diffraction at oblique angles of incidence.  Consideration of manufacturing cost and mechanical robustness of the two-layer AR coating design favors the largest possible pitch, as this choice minimizes the number of cuts required to cover a given area and makes the pillars larger and therefore stronger.  A 435 $\mu$m pitch leads to acceptable performance up to $\sim 175$ GHz  ($\sim 1.7$ mm) which is the upper edge of our band.

Comparing this relation between $v_f$ and $n_{eff}$ to analytic models \cite{Biber:2003, Motamedi} for the case where the electric field of the incident wave is perpendicular to the grooves, we found discrepancies that are reduced as the pitch decreases and we approach the quasi-static limit $p/\lambda << 0.1$. Given that our coating design does not operate in this limit we find the analytic models insufficiently accurate for our purposes.  Thus we resort to numerical simulation to optimize the geometry of our coating.

The variation in the electrical thickness compared to the physical thickness (Figure \ref{fig:designcurves2}, right) is shown for the case of a quarter wave AR coating with index $\sqrt {n_{Si}}$ .  Additional simulations show that the electrical thickness depends on the index of refraction of the material on either side of the metamaterial dielectric layer.  We account for this small effect in the final numerical optimization of the multilayer coating design.  

\begin{table}[b]
\caption{The parameters for the ACTPol AR coating.}
\begin{center}
\begin{tabular}{llllll}
\hline
parameter & symbol &  dimension &  dimension \\ 
                 &              &  ($\mu$m) &  (units of $p$) \\ \hline
pitch 	& $p$       	& $435$ 	& 	$1.0$\\ 
kerf{*}	& $k_1$   	& $41$ 	& 	$0.094$\\
		& $k_2$  	& $190$ 	& 	$0.437$\\
depth	& $d_1$ 	& $217$ 	& 	$0.499$\\
		& $d_2$ 	& $349$ 	& 	$0.802$\\
\hline		
volume fill factor{*} & $v_f$$_1$ 	& $0.820 $ & \\
		                       & $v_f$$_2$ 	& $0.317 $ & \\
\hline
\end{tabular}\\
\end{center}
{*} nominal average values\\

\label{tab:designparams}
\end{table}

\begin{figure}[t]
\centerline{\includegraphics[width=0.5\textwidth, keepaspectratio]{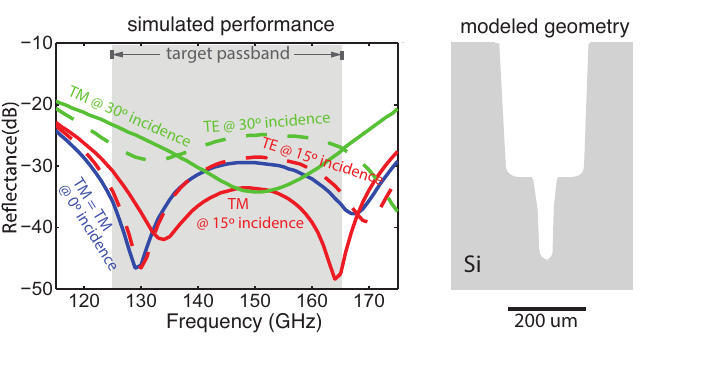}}
\caption{\footnotesize {\it Left:} A simulation of the performance of the AR coating designed for the ACTPol lenses at a range of angles of incidence. {\it Right:} The unit cell geometry (based on measured cut profiles) modeled using HFSS. }
\label{fig:ACTPoldesign}
\end{figure}

With these relations (Figure \ref{fig:designcurves2}) in hand we convert the analytic design of our two layer coating into parameters for the pillar geometry. In the final step we performed a numerical optimization of the coating at  $15^\circ$ angle of incidence using HFSS, with the constraint that the cut geometry must match the kerf geometry of commercially available dicing saw (See Section \ref{sec:descriptionofdicingsystem}). This step enables manufacturability, accounts for any discrepancies between the electrical and physical thickness implied by our numerical simulations, and improves the performance at larger angles of incidence. Table \ref{tab:designparams} gives the parameters for the resultant design and Figure~\ref{fig:ACTPoldesign} shows the simulated performance for this coating as a function of frequency for several ($0^\circ, 15^\circ$, and $30^\circ$) angles of incidence.  Even at $30^\circ$ incidence the band averaged reflections are at $-26$ dB and differences between the two linear polarization states (TE and TM, see  \cite{Jackson:1998} for definition) are below 0.5\% for all frequencies and angles.  At $15^\circ$ incidence, average reflections are below $-31$ dB. The tolerance of the design to various possible manufacturing errors are described in detail in Appendix A.

\section{ACTPol Lenses}
\label{sec:fab}

The ACTPol lenses are cylindrically symmetric plano-convex designs in which the convex surface is aspheric - a conic section with four perturbing terms proportional to the fourth, sixth, eighth, and tenth order of the distance from the axis. The design optimization procedure for the re-imaging optics is similar to that described in \cite{Fowler:2007}; the primary difference being that the ACTPol optics are required to be image-space telecentric to optimize the coupling to the flat feedhorn arrays.  This was accomplished by constraining the chief rays at each field point to be near normal incidence at the focal plane and by allowing the tilts of lens 2, lens 3, and the focal plane to vary during the optimization.  The resulting design achieves Strehl ratios greater than 0.93 across the 150 GHz focal planes without accounting for the Gaussian illumination of the feedhorns, which effectively improves the image quality. All three of the ACTPol optics tubes use the same three silicon lens designs with the positions and tilts adjusted to optimize the coupling to the ACT Gregorian telescope \cite{Fowler:2007}.

\begin{figure}[t]
\centerline{\includegraphics[width=0.28\textwidth, keepaspectratio]{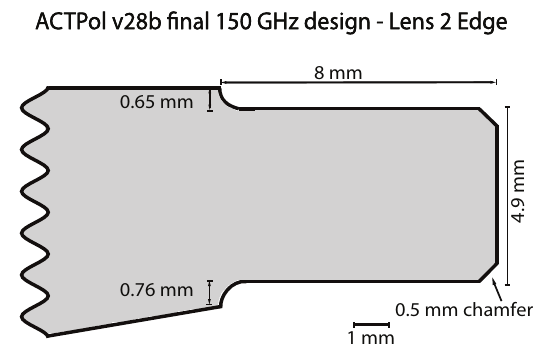}}
\caption{\footnotesize Cross-section of the perimeter of the ACTPol lens 2, which is designed to enable machining of the AR coating and clamping the lens without damaging the AR coating.  }
\label{fig:lens_perimeter}
\end{figure}

The AR coating approach we have developed constrains the perimeter of the lens designs to enable accurate clamping during both machining of the AR coatings and cryogenic cycling of the lenses.  Figure~\ref{fig:lens_perimeter} shows a cross-section of the perimeter of an ACTPol lens design.  Each lens includes a handling ring along its perimeter that is not AR coated and is used for mounting the lens during machining as well as    clamping to cool the lens in the cryogenic receiver. The outer corners of the perimeter are chamfered to minimize chipping.  There are steps rising from the perimeter to the lens surface to provide clearance between the perimeter clamping region and the AR coating.  The lens blanks were machined by Nu-Tek Precision Optical Corporation and achieved 5 $\mu$m tolerances.

\section{Properties of Silicon }
\label{sec:siliconprops}

Silicon manufacturing can produce different grades of material, such as ultra high purity silicon produced by the float zone process \cite{Duffar} that has a negligible density of impurities and silicon produced by the Czochralski process \cite{Duffar} that has a higher level of impurities. As charge carriers and associated states introduced by impurities are the cause of dielectric loss, use of the highest available purity (as inferred from the room temperature resistivity) silicon minimizes the dielectric losses. Measurements of the refractive index, dielectric permittivity and loss tangent of various high purity and high resistivity silicon sample over a range of frequencies and temperature have been reported \cite{Krupka:2006, Afsar:1994, Parshin:1995}. For example, ultra high purity silicon has been measured to have a loss tangent of  $\sim 1 \times 10^{-5}$ at room temperature which would correspond to $< 1\%$ loss in the ACTPol optical system. Unfortunately, surface tension limits the zone-melt purification technique used to produce ultra high purity silicon to diameters below about 200 mm. Therefore, the substrates available for large diameter lenses considered in this work must be fabricated from Czochralski silicon. For our low temperature application we expect the bulk conductivity of silicon to freeze out dramatically reducing the dielectric losses, however, other loss mechanisms can persist in the desired design band. We validate this general picture of the dielectric loss in silicon by optically characterizing the influence of the bulk resistivity as a function of frequency and sample temperature. 

Samples produced by the Czochralski process and readily available in 450 mm diameter stock with bulk resistivities specified to be in the range of 1 to $> 500~\Omega$-cm were characterized both at room temperature and 4K using a Bruker 125 high-resolution Fourier Transform Spectrometer (FTS) with a Oxford Cryostat CF continuous liquid helium flow sample chamber. The cryostat is equipped with pair of 75$~\mu$m thick polypropylene windows that enable spectral measurement while allowing the sample to be held at a regulated temperature. The samples were cut to have a typical thickness of $180~\mu$m, double side polished, and placed in a 25~mm diameter optical test fixture at the focus of an $f/6$ beam.  The reflections from the two surfaces of the sample form a Fabry-Perot resonator for which the modeling is relatively simple permitting measurement of dielectric properties. The silicon samples are boron doped (p-type) to adjust the resistivity. The $> 500~\Omega$-cm resistivity silicon used in the ACTPol lenses has the minimum dopant level.  

Each sample's transmission was measured between $240$ GHz ($8$ icm) and $18$ THz ($600$ icm) using different combinations of sources, beam splitters, and detectors for three frequency bands between 240-450GHz (8-15 icm), 450-2850 GHz (15-95 icm), and 2.85-18 THz (95-600 icm).  

The spectral resolution employed, 7.5 GHz (0.25 icm), fully resolves the sample's spectral features.  Sliding stages permit the sample or a reference clear aperture to be moved into the FTS beam while in the cryostat for {\it in situ} calibration. 

\begin{figure}[htbp]
	\centering
		\includegraphics[width=0.50\textwidth]{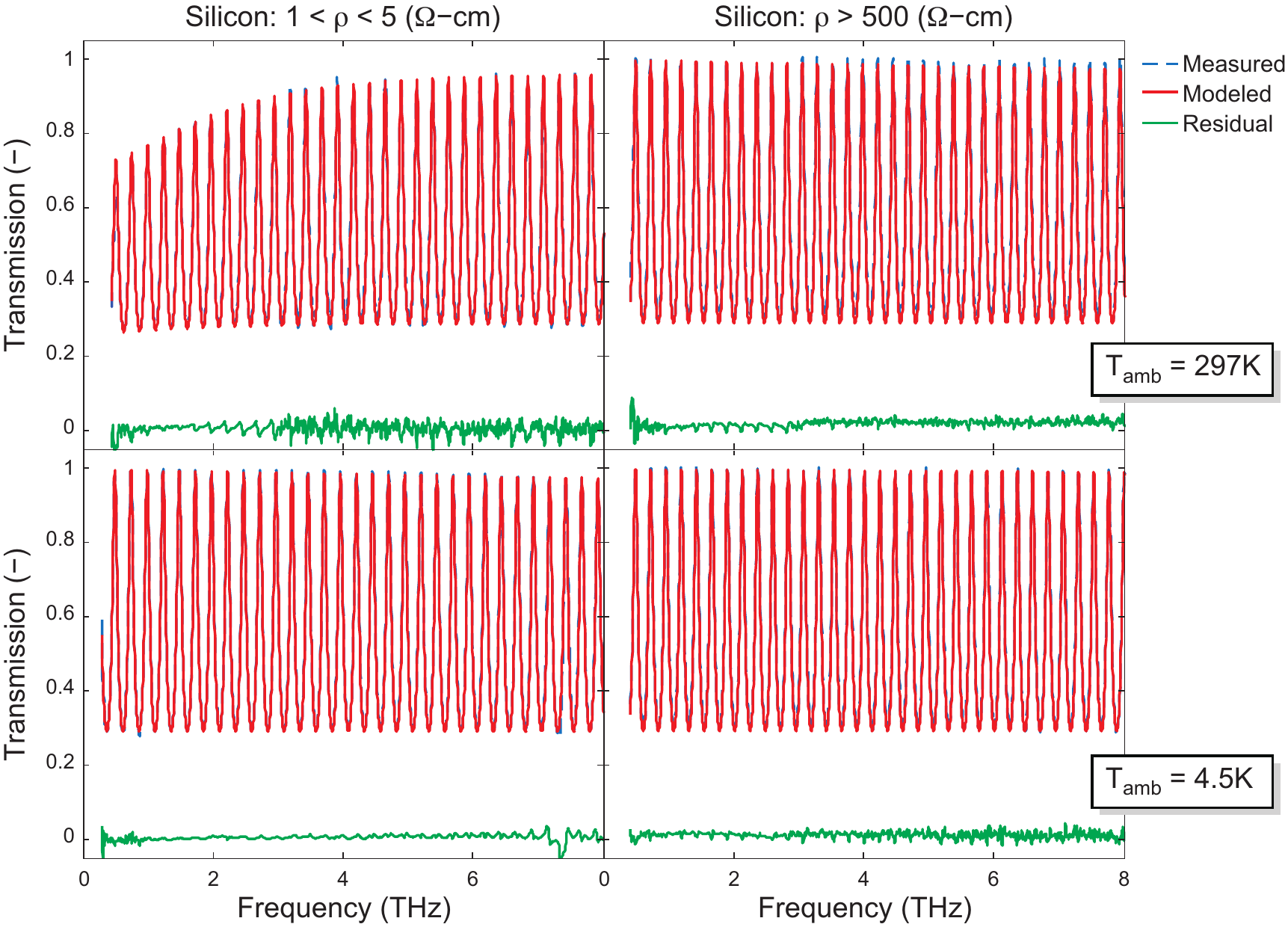}
	\caption{\footnotesize Measured silicon transmission at room-temperature (upper panels) and 4.5~K (lower panels). The left and right columns in the panel are for samples specified as $1<\rho<5~\Omega$-cm and $\rho>500~\Omega$-cm respectively, the sample thickness is $\sim180 \mu$m thick.  The figure contains the measured FTS data (red), model (dashed blue), and residual (green).}
	\label{fig:Silfex_residuals}
\end{figure}

The transmission spectrum was modeled as a series of homogeneous plane parallel dielectric layers~\cite{Yeh}.  The dielectric function for the silicon was approximated by a classical Drude dispersion model \cite{Button}:
\begin{equation}
	\varepsilon^*_r(\omega) = \varepsilon^*_\infty - \frac{\omega^2_{\mbox{{\tiny\it p}}}}{\omega\cdot(\omega+i\Gamma)}
\label{eq:Drude}
\end{equation}
where ${\varepsilon^*_r}=\varepsilon'_r+i\varepsilon''_r$ is a complex function of frequency $\omega$, the damping rate $\Gamma$, and the contribution to the relative permittivity $\varepsilon^*_\infty$ of higher energy transitions. The plasma frequency and the damping rates are related by, $\omega^2_{\mbox{{\tiny\it p}}} = \Gamma/ \varepsilon_o \rho$, where $\varepsilon_o $ is the permittivity of free space and $\rho$ is the sample's bulk resistivity~\cite{VanExter}. In this approximation, the material's free carriers are treated as classical point charges undergoing random collisions and the resulting damping is assumed to be independent of the carrier energy. We find this representation suitable to represent the sample's properties over the spectral and temperature ranges of interest.

The sample thickness is known to $\pm0.5\mu$m at room temperature and corrected for thermal contraction as a function of temperature~\cite{Okada}. Since the fringe rate is proportional to the product of the refractive index and sample thickness,uncertainties in the sample thickness directly limit the precision of the determination of $\Re \left( \varepsilon^*_\infty \right)$.  Measurements of an optically polished crystalline quartz sample with accurately known thickness were used to measure a 1\% amplitude uncertainty across the entire FTS band.  This calibration uncertainty leads to a corresponding reduction in the measurement's sensitivity to $\Im \left( \varepsilon^*_\infty \right)$.  These uncertainties were accounted for in fitting the FTS data.  These fits produced root-mean-square deviations in the range of $\sim$0.005-to-0.016 between the model and the observation spectra.  Representative data for the $\rho>500$ and $1<\rho<5~\Omega$-cm samples are shown in Figure \ref{fig:Silfex_residuals} for sample physical temperatures of $T_{\it phys}$=297 and 4.5~K.\\

\vspace{-0.2 in}
\begin{table}[t]
  \centering
  \caption{Silicon Drude Model Fit Parameters.}
    \begin{tabular}{llllll}\\ \hline
  $\rho(T_{\it phys})$      & $T_{\it phys}$  & $\varepsilon'_\infty$  & $\varepsilon''_\infty$  & $\Gamma/2\pi$  & $\omega_{\mbox{{\tiny\it p}}}/2\pi$  \\
 $[\Omega$-cm$]$ & [K]                   & $[-]$                          & $[-]$                          & $[$THz$]$                                       & $[$THz$]$         \\ \hline
  
  $>$500      &    297  &   11.7(1)   &             0.0015  &  $-$  &  $-$  \\
                    &     4.5  &   11.5(5)   &  $< $0.0008  &  $-$  &   $-$  \\ \hline
 
 1-to-5          &    297  &  11.6(55)  &             0.0046  &        0.571 &       1.60    \\
                    &    200  &  11.5(52)  &             0.0049  &        0.572 &        0.80   \\
                    &    100  &  11.4(78)  &             0.0028  &        0.459 &        0.26   \\
                    &      70  &  11.4(66)  &             0.0011  &        0.162 &        0.20   \\                       
                    &      30  &  11.4(64)  &  $< $0.0008  &  $-$ &  $-$     \\
                    &      10  &  11.4(62)  &  $< $0.0008  &  $-$ &  $-$     \\
                    &     4.5  &  11.4(62)  &  $< $0.0008  &  $-$ &  $-$     \\  \hline
  
    \end{tabular}
  \label{tab:results}
\end{table}

The results of these fits are shown in Table \ref{tab:results}. As anticipated the finite plasma frequency and damping rate contribute significantly to the room temperature losses, however, the $\rho>500~\Omega$-cm sample could be approximated by a dielectric constant over the spectral range of interest (e.g., $\Gamma$ and $\omega_p$ are consistent with zero).  The influence of free carrier collisions at room temperature on the optical response becomes more pronounced in the $1<\rho<5~\Omega$-cm sample, manifesting as a reduction in transmission at low frequencies. However, as the  $1<\rho<5~\Omega$-cm  sample is cooled, bulk conduction is suppressed and this effect shifts to lower frequencies than of interest for millimeter wave applications. The observed dielectric parameters are a weak function of temperature below $\sim30$~K as anticipated given exponential thermal dependance of the bulk resistivity {\cite{Shklovskii}}.

Both samples show $\sim 1 \%$ shifts in the magnitude of the index of refraction ($n \approx \sqrt {\epsilon_{\infty}'}$) upon cooling.  This shift is accounted for in our optical design.  At low temperature both the high and low resistivity sample give an upper limit for the loss tangent of  $\tan \delta \sim {\epsilon_\infty''} / {\epsilon_\infty'} < 7 \times 10^{-5}.$  For the ACTPol optics at 150 GHz, this corresponds to an absorptive loss of $<5\%$ for our optical system comprising three lenses. The ACTPol lenses were fabricated from $\rho > 500$ $\Omega$-cm silicon as it may reduce loss and facilitates room temperature testing.   Both high and low resistivity silicon would provide acceptable performance in 4K cryogenic applications, but given the small cost differential high resistivity silicon is the natural choice. Although we have used high-resistivity silicon, our models and observations suggest that the carriers will be frozen out at 4K for the range of bulk resistivities considered here. We note that the conductivity of silicon at room temperature, and therefore the loss, is a strong function of UV irradiance. We have not investigated this dependence cryogenically.\\

\section{Coating Fabrication}

The coatings were fabricated using silicon dicing saw blades which are available in widths ranging from 10 to hundreds of microns.  Mounted on a commercial silicon dicing saw machine these blades repeatedly cut with micron level precision and cutting speeds up to several centimeters per second.  Unfortunately, commercial silicon dicing machines are not designed for curved surfaces and cannot accommodate the large diameter lenses required for this project. Therefore we constructed a custom three axis silicon dicing system to fabricate these coatings.

\begin{figure}[t]
\centerline{\includegraphics{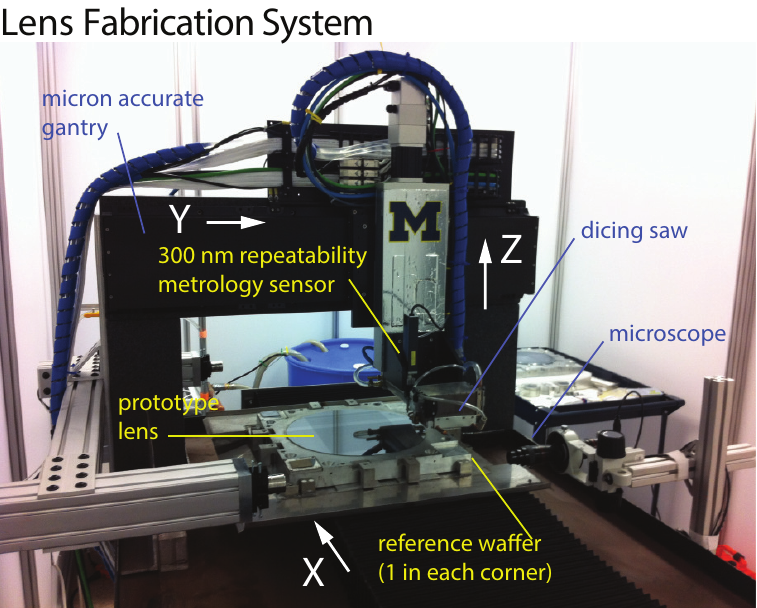}}
\caption{\footnotesize The custom, micron accurate, three axis silicon dicing system used to fabricate coatings on the lenses.  The labels identify the key components and axes described in the Section \ref{sec:descriptionofdicingsystem}. }
\label{fig:dicingsystem}
\end{figure}

\label{sec:descriptionofdicingsystem}

The fabrication system is shown in Figure \ref{fig:dicingsystem}.  It consists of a micron accurate three axis stage on which we mount an air bearing dicing spindle and a micron accurate depth gauge.  The spindle and retractable depth gauge are attached to the vertical stage (Z-axis) which rides on the horizontal (Y-axis) stage.  A lens is mounted on an aluminum mounting plate on a horizontal stage (X-axis) below the spindle.  This plate permits the lens to be rotated by $0^\circ$, $90^\circ$, $180^\circ$, and $270^\circ$ and carries a reference wafer that is used in setting up the blades prior to cuts. A side looking microscope is mounted parallel to the Y-axis to characterize test cuts on the reference wafer.  Flood cooling water is sprayed on the dicing blade while cuts are made to carry away debris.  A temperature controlled water bath is used to regulate the temperature of the spindle and maintain the flood coolant and air surrounding the dicing system within $1^\circ$ C.

Our system does not show any appreciable change in cut shape or surface damage when cutting at the maximum travel speed of 50 mm/s.  We conservatively operate at 25 mm/s for which it takes a total of 12 hours of machining per lens side.  Factoring in setup time it takes 6 8-hour days to fabricate a single lens.

\begin{figure}[t]
\centerline{\includegraphics{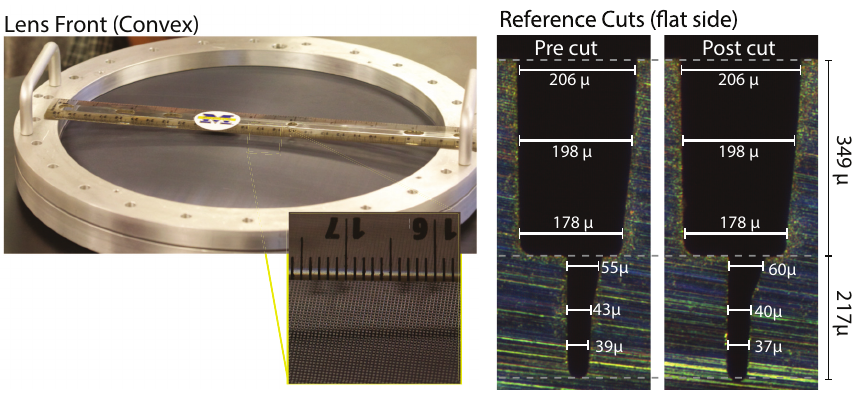}}
\caption{\footnotesize {\it Left:} Photograph of the curved front surface of an AR coated ACTPol silicon lens (lens 2) with zoomed view of a small patch. {\it Right:} Photographs taken with side looking metrology microscope of a reference wafer taken prior to and after cutting the coating on a lens.}
\label{fig:lensfabed}
\label{fig:metrology}
\end{figure}

Figure~\ref{fig:lensfabed} shows photographs of one of the ACTPol AR coated silicon lenses. The fabricated coating is sufficiently robust to permit handling the lenses by touching the AR coatings.  The manufacture resulted in less than 10 out of 500,000 posts with damage visible by eye.  The shape of the cut profiles was evaluated by making cuts on a reference silicon wafer before cutting (pre-cut) and after cutting (post-cut) the lens and measuring them using the side looking microscope.  Comparison of the pre-cut and post-cut measurements (See Figure \ref{fig:lensfabed}) show that the wide blade cuts repeatably with negligible evolution to cut profile while the the narrow blade shows some evolution in width in the upper third of the cut.  Simulations show that this evolution, leads to a few tenths of a percent increase in the reflectance.  The accuracy of the depths of the cuts is limited by a 3 micron uncertainty in the zero point for the blade depth and a $\pm 7$ micron quadruple  warp in the lens. This warp is due to imperfections in the lens mounting plate that stress the lens.  

\begin{figure}[t]
\centerline{\includegraphics{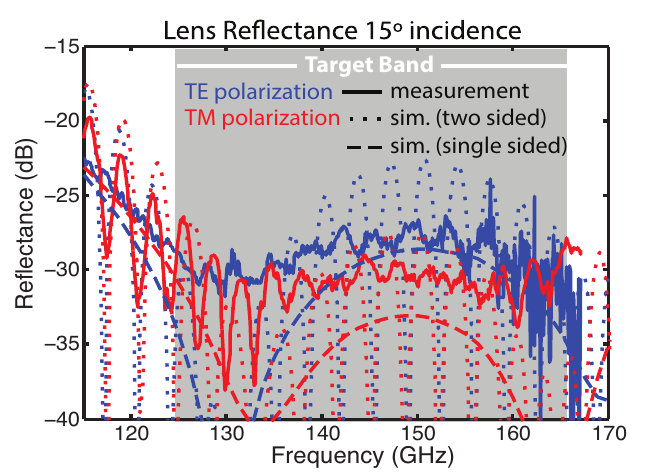}}
\caption{\footnotesize Comparison between simulations and measurements of the fabricated lens. The two sided simulations model the AR coating on both sides of a flat silicon sample of thickness equal to that of the lens at its center. The single sided simulations model the AR coating on only one side of a flat silicon sample.}
\label{fig:prototypeprform}
\end{figure}

The reflectance of the completed ACTPol lenses were measured using a scalar reflectometer.  This reflectometer consisted of a tunable narrow band continuous wave source which illuminated the flat side of lens through a $20^\circ$ full width half power horn, from a distance of 5 cm, and tilted at an angle of $15^\circ$ relative to the lens surface normal.  A receiver consisting of an identical horn coupled to a detector diode was placed at the mirror image of the source horn relative to the plane defined by the point of maximum illumination of the lens and perpendicular to a line joining the source and receiver.   This system was calibrated by (1) placing an aluminum reflector at the same position as the lens to normalize the peak reflection to unity, and (2) by removing the lens and calibration reflector to measure the stray reflections which were found to be negligible.  

The results of this reflectance measurement are shown in Figure \ref{fig:prototypeprform} for both the TE and TM polarizations. Precise modeling of this measurement configuration requires accounting for the interference between reflections from the flat and curved lens surface, which is beyond the scope of this work.   However, the results are expected to be intermediate between the reflectance from a single AR coated surface (single sided) and the easily simulated result for two flat surfaces separated by the central thickness of the lens (two sided).  These two cases are presented in Figure \ref{fig:prototypeprform} for both linear polarization states.  These simulations incorporate the measured cut profiles shown in Figure \ref{fig:lensfabed}.  We empirically found that moving the lens so that it is well centered on the beam increases the interference effects (TM case) while moving the lens off center reduces these effects (TE case).  Based on the reasonable agreement between these simplistic simulations and the measurement we are confident we are reducing reflections to a few tenths of a percent over the range of angles of incidence required for ACTPol.

\section{Conclusion}

We have described a new approach for AR coating silicon lenses over broad bandwidths and a range of angles of incidence.  Simulations, backed up by measurements of an AR coated lens, show that the fabricated coating of the lenses presented here can reduce reflections below few tenths of a percent between 125 and 165 GHz for angles of incidence between 0 and $30^\circ$ for cryogenic applications. We have developed a micron-accurate 3-axis silicon dicing saw facility and are using it to manufacture AR coated lenses for ACTPol. We have also shown that a range of p-type silicon doping levels can achieve low loss at cryogenic temperatures using silicon samples from boules as large as 45 cm diameter. This approach for implementing wide-bandwidth AR coated silicon lenses is applicable for millimeter and sub-millimeter wavelength ranges and can be expanded to wider bandwidth by adding additional layers to the AR coating.

\section*{Acknowledgement}

This work was supported by the U.S. National Science Foundation through awards AST-0965625 and PHY-1214379 and NASA through the NASA Space Technology Research Fellowship training grant NNX12AM32H. The authors would like to thank Ki Won Yoon and Molly Dee for useful discussions.

\begin{appendix}

\section{Tolerances}

\begin{figure}[b]
\centerline{\includegraphics{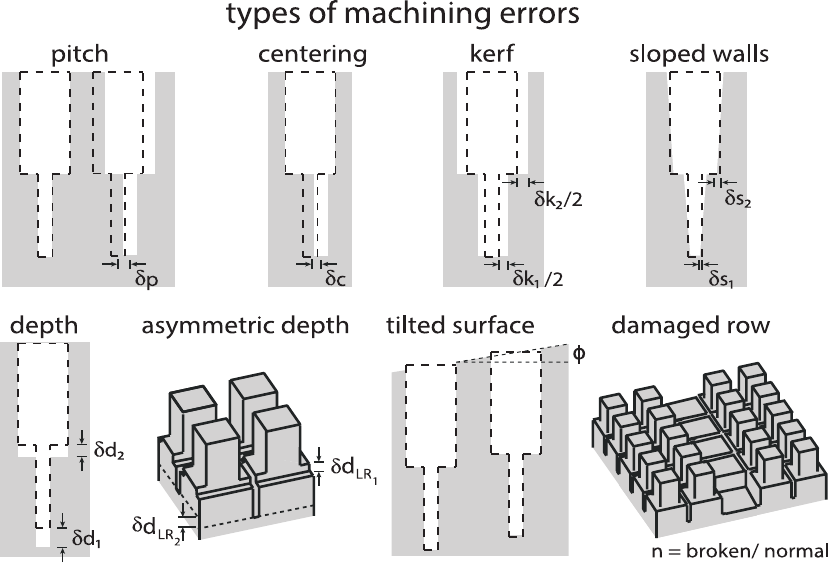}}
\caption{\footnotesize This figure shows the families of machining errors considered in the tolerance analysis and described in the text.  The dashed lines show the target shape for the grooves and that for the pillars while the white space in the grey represents the material as actually cut. Therefore difference between the dashed region and the white region represent machining errors.  For the asymmetric depth and damaged row errors  the sketches show the geometry of the errors. }
\label{fig:machinetol}
\end{figure}

The manufacturing tolerances for the eight types of machining errors shown in Figure~\ref{fig:machinetol} were evaluated based on a fiducial design. These include: (1) errors in the pitch, $\delta p$; (2) errors in the centering of the deeper groove relative to the shallow groove, $\delta c$; (3) errors in the kerf widths,  $\delta k_1$, $\delta k_2$; (4) slopes in the walls of the grooves parameterized by  $\delta_{S1}$ and $\delta_{S2}$ which represent the distance the upper and lower groove tilt inward at the groove bottom if the mean width is fixed; (5) errors in the depth of either grooves  $\delta d_1$, $\delta d_2$, (6) differences in the depth of grooves in the two orthogonal directions  assuming the correct mean depth are parameterized $\delta d_{LR_1}$ for the upper and $\delta d_{LR_2}$ for the lower groove; (7) application of this coating to a surface tilted at an angle $\phi$ relative to the bottom edge of the saw and where the depth of each groove is measured at the center of the groove (e.g., the effect of applying this coating at the edge of a curved lens);  and (8) rows of broken posts. The tolerance to variation in the refractive index of the silicon substrate, $\delta n_{Si}$ was also evaluated. The tolerances for these parameters were quantified based on a fiducial model similar to what was fabricated but with straight walled pillars using HFSS simulations in which the parameters in each family were varied separately. For the majority of these effects the results were distilled to a band averaged reflection at $15^\circ$ angles of incidence. The sensitivity to each parameter was quantified as the displacement needed to bring about a $3$ dB increase in reflectance.  For the case of tilted surface and for broken posts the results of a small number of simulations were evaluated and compared to the fiducial performance.  In the remainder of this section we discuss these results.

\begin{figure}[t]
\centerline{\includegraphics{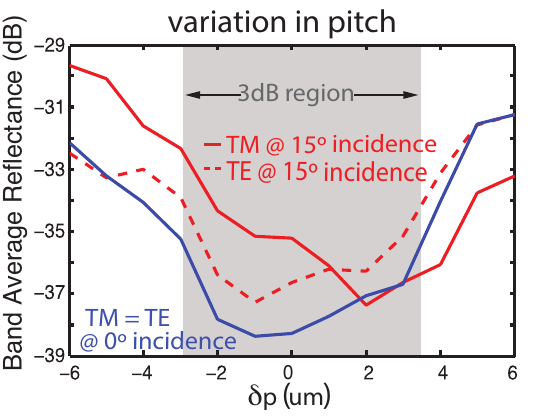}}
\caption{\footnotesize The band averaged reflectance at a $15^\circ$ angle of incidence as a function of changes in the pitch.  The grey region shows range for which the performance is within 3dB of the fiducial design. }
\label{fig:pitchtol}
\end{figure}

\paragraph{Pitch:}
Variations in the pitch change the effective index of both layers.   Figure~\ref{fig:pitchtol} shows the band averaged reflectance as function of variation in the pitch.  A conservative tolerance of $\delta p < 3 \mu$m ($\delta p/p < 0.0075$) limits the change in reflectance to be less than 3 dB.

\paragraph{Centering:}  The performance was insensitive to errors in the centering of the deeper grooves within the shallower grooves.   This is consistent with both layers of pillars behaving as layers with a constant effective index of refraction.

\begin{figure}[t]
\centerline{\includegraphics{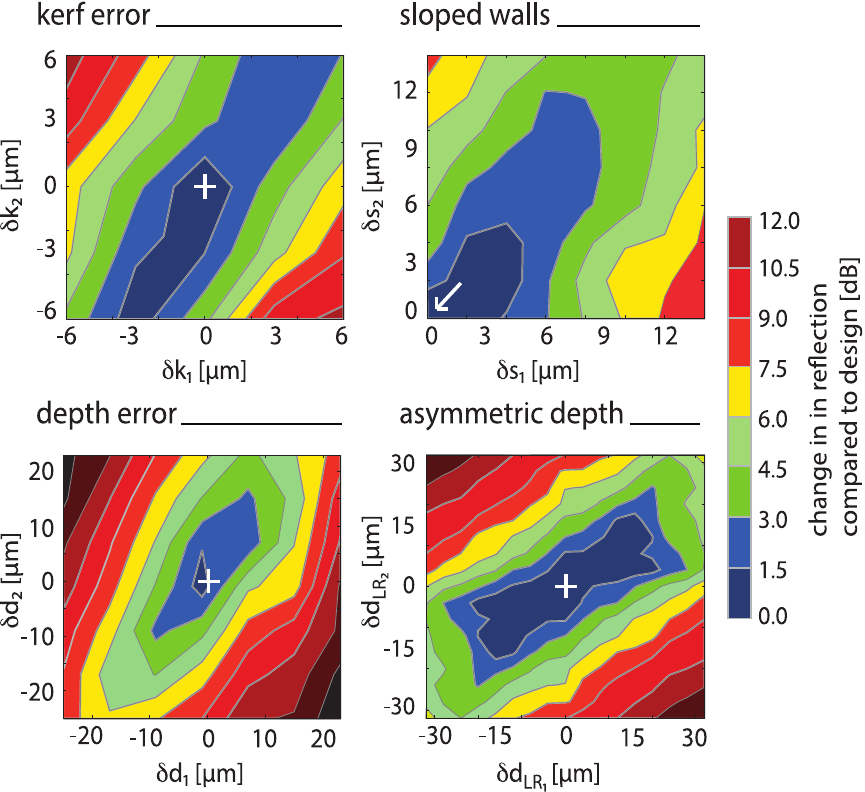}}
\caption{\footnotesize This figure shows the impact of errors in the kerf, slopes in the sidewalls, overall depth errors, and asymmetric differences in depth between the two orthogonal cuts.  The color scale presents the degradation in reflection at $15^\circ$ incidence compared to the simulations in Figure~\ref{fig:ACTPoldesign}.  The fiducial design is at the coordinate (0,0) in all the plots as highlighted by a ``+'' or arrow.  The outer edge of the light blue contour (see label) represents 3dB degradation.  The horizontal (vertical) axis represents errors in the inner (outer) layer of the coating for the labeled parameters.  }
\label{fig:contoursysts}
\end{figure}

\paragraph{Kerf:}  The upper left panel of Figure~\ref{fig:contoursysts} presents the impact of errors in the kerf.  Controlling $\delta k_1$ and $\delta k_2$ to 3 $\mu$m insures a less than 3dB degradation in performance. Varying both parameters in the same direction (e.g., $\delta k_1 = \delta k_2$) has little impact on the performance.  We have chosen the fiducial design to be near the high end of the most favorable region to make the manufacture relatively immune to blade wear, which could have (but did not) narrow the kerf width as machining proceeds.

\paragraph{Sloped Walls:}   The upper right panel of Figure~\ref{fig:contoursysts} presents the impact of slopes in the walls left behind by the dicing saw.  Conservative estimates for the $3$ dB tolerance for these two parameters are 6 $\mu$m for $\delta s_1$ and 3 $\mu$m for $\delta s_2$.  Sloping both layers by a similar amount results in negligible degradation.

\paragraph{Depth:} The lower left panel of Figure~\ref{fig:contoursysts} presents the impact of depth errors.  The $3$ dB tolerances for these parameters are 7.5 $\mu$m for $\delta d_1$ and 10 $\mu$m for $\delta d_2$.  Expressed in terms of optical path length the sensitivities are identical.

\paragraph{Asymmetric Depth:}  The lower right panel of Figure~\ref{fig:contoursysts} presents the impact of asymmetric depth.  The $3$ dB tolerances for these parameters are 15 $\mu$m for $\delta d_{LR_1}$ and 12 $\mu$m for $\delta d_{LR_2}$ for asymmetry in directions common to both grooves. This error also affects the $90^\circ$ rotational symmetry of the geometry resulting in birefringence but the cross-polarization can be no larger than the absolute reflection.

\begin{figure}[t]
\centerline{\includegraphics{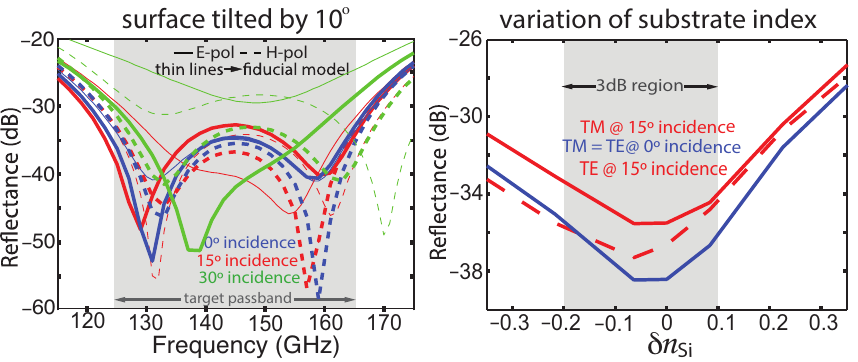}}
\caption{\footnotesize {\it Left:} The simulated impact of having the surface of the lens tilted by $10^\circ$.  {\it Right:} The simulated impact of variation in the refractive index of the silicon substrate.}
\label{fig:tilt_broke}
\end{figure}

\paragraph{Application to tilted surfaces:} The left panel of {Figure~\ref{fig:tilt_broke}, left} compares the performance of this coating applied to a surface tilted by $10^\circ$ compared with the performance when applied to a flat surface.  Fortuitously, this effect improves performance at large angles of incidence and has little effect at small angles.  Applying this coating by making grooves parallel to the lens symmetry axis of a curved lens will produce acceptable performance.

\paragraph{Groups of broken posts:}  Assessing the impact of the broken posts is not straight forward because the region and extent of illumination varies between the three lenses. The degradation due to broken posts would thus be a function of the effective aperture area of the lenses and the location of the defects in addition to the total area affected. Accurately quantifying this is beyond the scope of this work. Qualitatively, the degradation would be negligible if the fraction of total area affected is small and more importantly, the length scale of the individual affected areas is smaller than the wavelength of incident light. Simulating the impact of broken posts is computationally difficult.  Therefore we resorted to a simplified model consisting of a unit-cell containing seven posts in a row with one broken off completely.  This is equivalent to having one seventh of the posts destroyed.  This pessimistic case produces band averaged reflections below 2\% which is acceptable though it does produce non-trivial cross-polarization. We used simulation to estimate that  keeping the number of broken posts below 1/700 would result in a negligible degradation in the overall performance of the coating.

\begin{table}[b]
\caption{\footnotesize A summary of the 3 dB sensitivities for all parameters described in the text. }
\begin{center}
\begin{tabular}{cccccc}
\hline
machining error 	&parameter 	&  sensitivity	 	& tolerance \\
				&		 	&  $\mu$m 		& (units of $p$)\\
\hline
pitch ($p$)  		& $\delta p$ 	& $3$ & $7.5 \times 10^{-3} $\\
centering	 		& $\delta c$ 	& see text & \\
kerf	 			& $\delta k_1$ 	&  $3$ & $7.5 \times 10^{-3} $\\
	 			& $\delta k_2$ 	& $3$ & $7.5 \times 10^{-3} $\\
sloped walls		& $\delta s_1$ 	&  $6$ & $1.5 \times 10^{-2} $\\
	 			& $\delta s_2$ 	& $3$ & $7.5 \times 10^{-3} $\\
depth			& $\delta d_1$	& $7.5$ & $1.9 \times 10^{-2} $\\
	 			& $\delta d_2$ 	& $10$ & $2.5 \times 10^{-2} $\\		
asymmetric depth	& $\delta d_{LR_1}$ & $15$ & $3.8 \times 10^{-2} $\\
	 			& $\delta d_{LR_2}$ & $12$ & $3 \times 10^{-2} $\\	\hline				
tilted surface		& $\phi$ 		&see text&\\
damaged row		& $r$ 		& $\sim 1/700$ & NA\\
index 			& $\delta n$ 	& $0.1$ & NA\\
\hline			
\end{tabular}\\
\end{center}

\label{tab:tols}
\end{table}

\paragraph{Index:} {Figure~\ref{fig:tilt_broke}, right} shows the band averaged reflectance as a function of variation in the index.  A conservative tolerance for the index is $\delta n_{Si} < 0.1$ ($\delta n_{Si}/n_{Si} < 0.03$). This leads to increases in the reflectance of less than 3 dB over the fiducial design.  This tolerance is weaker than that imposed by the optical design.\\

The 3dB sensitivies derived in this section are summarized in Table~\ref{tab:tols}. The band averaged reflectivity of the fiducial design is -31 dB at $15^\circ$ incidence which is significantly better than the requirements for the system.

\end{appendix}


\begin{thebibliography}{99}

\bibitem{Bintley:2012}
D. Bintley et al., 
``{Commissioning SCUBA-2 at JCMT and Optimising the Performance of the Superconducting TES Arrays}", Journal of Low Temperature Physics, {\bf 167}, 152-60 (2012).
\bibitem{Niemack:2008}
M.D. Niemack et al., 
``{A Kilopixel Array of TES Bolometers for ACT: Development, Testing, and First Light}", Journal of Low Temperature Physics, {\bf 151}, 690-6 (2008).
\bibitem{Padin:2008}
S. Padin et al., 
``{South Pole Telescope optics}", Applied Optics {\bf 47}, 4418-28 (2008).
\bibitem{Hanany:2012}
S. Hanany, M. D. Niemack, and L. Page, ``{CMB Telescopes and Optical Systems}", Planets, Stars and Stellar Systems, Volume 1: Telescopes and Instrumentation (arXiv:1206.2402, in press).
\bibitem{Thompson:1961}
J. C. Thompson and B. A. Younglove, ``{Thermal Conductivity of Silicon at Low Temperatures}", J. Phys. Chem. Solids, Pergamon Press 1961. Vol. 20, Nos. 1/2, pp. 146-149.
\bibitem{Sciver:2012}
S. W. Va Sciver, Helium Cryogenics, \emph{International Cryogenics Monograph Series}, Springer Science+Business Media, LLC 2012, ch. 2.
\bibitem{Collin:1990}
R.E. Collin, \emph{Field Theory of Guided Waves}, 1990, McGraw-Hill, McGraw-Hill, New York, pp. 749-786. 
\bibitem{Smith:2006}
D.R. Smith and J.B. Pendry, ``Homogenization of metamaterials by field averaging (invited paper)", March 2006, Journal of the Optical Society of America B, Vol. 23, No. 3, pp. 391-403.  
\bibitem{Kildal:1984}
P.-S. Kildal, K. Jakobsen, and K. Sudhakar Rao, ``Meniscus-lens-corrected corrugated horn: a compact feed for a Cassegrain antenna," IEE Proc. {\bf 131}, 390-4 (1984).
\bibitem{Cohn:1961}
Cohn S. B., ``{Lens Type Radiators: Antenna Engineering Handbook}'', 1961 (McGraw-Hill, N.Y.)
\bibitem{Lau:2006}
J. Lau et al., 
``{Millimeter-wave antireflection coating for cryogenic silicon lenses}", Applied Optics, {\bf 45}, 3746-51 (2006).
\bibitem{Chapham}
P.B. Clapham and M.C. Hutley, ``Reduction of Lens Reflection by `Moth Eye' Principle", 1973, Nature, Vol. 244,  pp. 281-282.
\bibitem{Thornton}	
B.S. Thornton, ``Limit of the moth's eye principle and other impedance-matching corrugations for solar-absorber design", 1975, Journal of the Optical Society of America, Vol. 65, No. 3, pp. 267-270.
\bibitem{Motamedi}			
M.E. Motamedi, W.H. Southwell, and W.J. Gunning, ``Antireflection surfaces in silicon using binary optics technology", 1992, Applied Optics, Vol. 31, No. 22, pp. 4371-4376.	 
\bibitem{Zhang:2009}
J. Zhang, et al., 
``New artificial dielectric metamaterial and its application as a terahertz antireflection coating", Applied Optics, {\bf 48}, 6635 (2009).
\bibitem{Schuster:2005}
K.-F. Schuster, et al. 
``Micro-machined Quasi-Optical Elements for THz Applications", Sixteenth International Symposium on Space Terahertz Technology, held May 2-4, 2005 at Chalmers University of Technology. Gšteborg, Sweden., p. 524-528.
\bibitem{Han:2010}
P. Han et al., 
``Application of Silicon Micropyramid Structures for Antireflection of Terahertz Waves", IEEE Journal of Selected Topics in Quantum Electronics, Vol. 16, No. 1, January/February 2010
\bibitem{Niemack:2010}
M.D. Niemack, et al., ``ACTPol: a polarization sensitive receiver for the Atacama Cosmology Telescope'', Proc. SPIE {\bf 7741} (2010), arXiv:1006.5049.
\bibitem{Fowler:2007}
J.W. Fowler, et al., 
``{Optical design of the Atacama Cosmology Telescope and the Millimeter Bolometric Array Camera}", Applied Optics {\bf 46}, 3444-54 (2007).
\bibitem{McMahon:2012}
J.J. McMahon, et al., 
``Multi-chroic Feed-Horn Coupled TES Polarimeters", Journal of Low Temperature Physics, {\bf 167}, 879-84 (2012).
\bibitem{Shimon:2008}
M. Shimon et al., 
``{CMB Polarization Systematics due to Beam Asymmetry: Impact on Inflationary Science}", Phys. Rev. {\bf D77:083003} (2008).
\bibitem{Mackay:1989}
A. MacKay, ``Proof of Polarization Independence and Nonexistence of Crosspolar Terms for Targets Presenting n-Fold (n $>$ 2) Rotational Symmetry with Special Reference to Frequency-Selective Surfaces,Ó 1989, Electron.
Lett., vol. 25, no. 24, pp. 1624-1625.
\bibitem{Rytov:1956}
S. Rytov, ``The electromagnetic properties of finely layered medium", Soviet Physics JETP 2 (1956) 466-475.
\bibitem{Biber:2003}
S. Biber, et al., 
``Design of Artificial Dielectrics for Anti-Reflection-Coatings", 33rd European Microwave Conference, Munich (2003).
\bibitem{Brundrett:1994}
D.L. Brundrett, E.N. Glytsis, and T.K. Gaylord, ``Homogeneous layer models for high-spatial-frequency dielectric surface-relief gratings: conical diffraction and antireflection designs", 1 May 1994, Applied Optics, Vol. 33, No. 13, pp. 2695-2706.
\bibitem{Gaylord:1986}
T.K. Gaylord, W.E. Baird, and M.G. Moharam, ``Zero-reflectivity high spatial-frequency rectangular-groove dielectric surface-relief gratings", 15 December 1986, Applied Optics, Vol. 25, No. 24, pp. 4562-4567.              
\bibitem{Aspnes2:1982}
D.E. Aspnes, ``Local-field effects and effective-medium theory: A microscopic perspective", Am. J. Phys. 50, 704 (1982).
\bibitem{Aspnes1:1982}
D.E. Aspnes, ``Bounds on Allowed Values of the Effective Dielectric Function of Two-Component Composites at Finite FrequenciesÓ, 1982, Physical Review B, Vol. 25, No. 2, pp. 1358-1361.
\bibitem{Egan:1982}
W.G. Egan and D.E. Aspnes, ``Finite-wavelength effects in composite media", 1982, PRB, Vol. 26, No. 10, pp. 5313-5321.
\bibitem{Gentner:2006}
A. Wagner-Gentner, et al. 
``Low loss THz window", Infrared Physics and Technology 48 (2006) 249-253
\bibitem{Matthaei}
G. Matthaei, L. Young, E.M.T. Jones, \emph{Microwave Filters, Impedance-Matching Networks and Coupling Structures}, 1964, McGraw-Hill, New York, pp. 300-304.
\bibitem{Raguin:1993}
D.H. Raguin and G.M. Morris, ``Analysis of antireflection-structured surfaces with continuous one-dimensional surfaces profiles", 10 May 1993, Applied Optics, Vol. 32, No. 14, pp. 2582-2598.
\bibitem{Jackson:1998}
Jackson J.D., \emph{Classical Electrodynamics}, 1998, John Wiley \& Sons Inc.
\bibitem{Ansoft:HFSS}
``{Ansoft High Frequency Structure Simulator (HFSS) software package}", \url{http://www.ansys.com/Products/Simulation+Technology/Electromagnetics/High-Performance+Electronic+Design/ANSYS+HFSS}
 \bibitem{Duffar}
T. Duffar, \emph{Crystal Growth Processes Based on Capillarity: Czochralski, Floating Zone, Shaping and Crucible Techniques}, (Wiley, Blackwell, 2010).
\bibitem{Krupka:2006}
J. Krupka, et al., ``Measurements of Permittivity, Dielectric Loss Tangent, and Resistivity of Float-Zone Silicon at Microwave Frequencies", 2006, IEEE Transaction on Microwave Theory and Techniques, vol. 54, No. 11, pp. 3995-4000. 
\bibitem{Afsar:1994}
M.N. Afsar and H. Chi, ``Millimeter wave complex refractive index, complex dielectric constant, and loss tangent of extra high purity and compensated silicon", 1994, International Journal of Infrared and Millimeter Waves, vol. 15, pp. 1181-1188. 
\bibitem{Parshin:1995}
V. V. Parshin, et al. 
``Silicon as an Advanced Window Material for High Power Gyrotrons", 1995, Int. Journal of Infrared and Millimeter Waves, vol. 16, no. 5, pp. 863-877.
\bibitem{Yeh}
P. Yeh, \emph{Optical Waves in Layered Media} (Wiley, New York, 1988).
\bibitem{Button}
F. Gervais, ``High-Temperature Infrared Reflectivity Spectroscopy by Scanning Interferometry'' in \emph{Electromagnetic Waves in Matter}, Part I, Vol. 8 (Infrared and Millimeter Waves), K. J. Button, eds. (Academic Press, London, 1983), pp. 284--287.
\bibitem{VanExter}
M. Van Exter and D. Grischkowsky, ``Optical and Electronic Properties of Doped Silicon from 0.1 to 2 THz", 1990, Applied Physics Letters, Vol. 56, No. 17, pp. 1694-1696.
\bibitem{Okada}
Y. Okada and Y. Tokumaru, ``Precise determination of lattice parameter and thermal expansion coefficient of silicon between 300 and 1500K", 1984, J. Appl. Phys., Vol. 56, No. 2, pp. 314-320.
\bibitem{Shklovskii}
B.I. Shklovskii and A.L. Efros, \emph{Electronic Properties of Doped Semiconductors}, 1984, Berlin, Germany, Springer, ch. 4.

\end{thebibliography}
\end{document}